\documentclass[12pt]{article}

\usepackage{fancybox}

\usepackage{cite}
\usepackage{float}
\usepackage{amsfonts}
\usepackage{amsmath}
\usepackage{amsbsy}
\usepackage{graphicx}
\usepackage{amssymb}
\usepackage{amsthm}
\usepackage{bm}
\usepackage{epsfig}
\usepackage{latexsym}
\usepackage{pdflscape}
\usepackage{color}
\usepackage{here}
\usepackage{graphicx}
\numberwithin{equation}{section}

\DeclareMathOperator{\tr}{tr}

\begin{document}

\renewcommand{\thefootnote}{\fnsymbol{footnote}}

\begin{titlepage}
\begin{flushright}
{\footnotesize OCU-PHYS 446}
\end{flushright}
\bigskip
\begin{center}
{\LARGE\bf Prospects of the Nambu Bracket}\\
\bigskip\bigskip
{\large 
Kazuki Kiyoshige\footnote{\tt kiyoshig@sci.osaka-cu.ac.jp},
Sanefumi Moriyama\footnote{\tt moriyama@sci.osaka-cu.ac.jp},
and
Katsuya Yano\footnote{\tt yanok@sci.osaka-cu.ac.jp}
}\\
\bigskip
{\it Department of Physics, Graduate School of Science,
Osaka City University}\\
{\it 3-3-138 Sugimoto, Sumiyoshi, Osaka 558-8585, Japan}
\end{center}

\bigskip

\begin{abstract}
We review recent progress in formulating the worldvolume theory of M2-branes using the Nambu bracket.
Although it is generally agreed that this formulation should be replaced by another using the superconformal Chern-Simons theory, we try to pursue a possible complementary role played by the Nambu bracket.
Since the partition function of the superconformal Chern-Simons theory implies a hidden super gauge group, we study the supersymmetric generalization of the Nambu bracket.
We explicitly construct the superunitary Nambu algebra.

\end{abstract}
\end{titlepage}

\tableofcontents

\renewcommand{\thefootnote}{\arabic{footnote}}
\setcounter{footnote}{0}

\section{Introduction}\label{introduction}

In 1973, Nambu proposed a generalization of the Hamiltonian mechanics \cite{Nambu}.
In this dynamics, there appear multiple Hamiltonians $H$ and $K$ governing the time evolution
\begin{align}
\frac{d}{dt}f=\{f,H,K\}_\text{Nambu},
\end{align}
where $\{*,*,*\}_\text{Nambu}$ is called the Nambu bracket:
\begin{align}
\{f,g,h\}_\text{Nambu}=\varepsilon_{ijk}\,
\partial_if\,\partial_jg\,\partial_kh.
\label{NB}
\end{align}
Just as the Poisson bracket satisfies the Jacobi identity, this bracket also satisfies an identity, which is now called the fundamental identity or the Filippov identity.
Since the bracket was further generalized to those with $n$ entries, this algebra with three entries is called the three-algebra.

Recently, a proposal has been put forward to formulate the worldvolume theory of multiple M2-branes using the Nambu bracket.
This theory is called BLG theory after its founders, Bagger, Lambert, and Gustavsson \cite{BL,G}.
See Ref.~\cite{BLMP} for an extensive review of the BLG theory.
It is surprising to see Nambu's idea revived after three decades at the frontier of string theory.

The BLG theory is a 2+1D theory whose Lagrangian is schematically
\begin{align}
S_\text{BLG}
=\int d^3x\Bigl((DX)^2+\Psi D\Psi+\Psi[X,X,\Psi]+[X,X,X]^2\Bigr)
+S_\text{CS},
\end{align}
where the Chern-Simons term is given by
\begin{align}
S_\text{CS}=\frac{k}{4\pi}\int A\wedge dA+\frac{2}{3}A\wedge A\wedge A,
\end{align}
with the Chern-Simons level $k$.
Using the fundamental identity of the Nambu bracket, the gauge symmetry and the ${\mathcal N}=8$ supersymmetry were proved \cite{BLsym}.
With the correct symmetry and matter contents, it was expected that this theory would describe the worldvolume theory of M2-branes.
M2-branes are the fundamental 2+1D excitations in M-theory, which is a 10+1D theory proposed to unify five perturbative vacua of string theory.

However, it was found that the fundamental identity is very restrictive.
In fact, if we require the algebra to be finite-dimensional, positive-definite, and completely antisymmetric, it can be proved that there is essentially only one algebra \cite{Pap,GG}.\footnote{See Refs.~\cite{Gloop,BLS,HHM} for earlier studies on the fundamental identity and these requirements.}
This is in contrast to our expectation of describing $N$ multiple M2-branes.
So there were some trials to relax these requirements.

In the meantime, it was gradually realized that multiple M2-branes can be described by the supersymmetric generalization of the Chern-Simons theory called ABJM (Aharony-Bergman-Jafferis-Maldacena) theory \cite{ABJM,HLLLP2,ABJ}, following an early attempt in Ref.~\cite{S}.
In fact, among others, the free energy of this theory reproduces \cite{DMP1} the $N^{3/2}$ behavior\footnote{Moreover, all of the perturbative \cite{DMP2,FHM} and nonperturbative \cite{DMP2,MaPu,HMO2,CM,HMO3,HMMO} corrections are 
determined.
See Ref.~\cite{PTEP} for a review.}
of the degrees of freedom of the M2-branes \cite{KT}.
This is strong nontrivial evidence for the theory.

All of the studies on the supersymmetry, the large-$N$ behavior, and the moduli space indicate that the ABJM theory in the infrared limit describes the multiple M2-brane system correctly.
However, at the same time, it is known that for the flat case $k=1$ not all of the supersymmetries ${\cal N}=8$ are realized explicitly.
This leads us to expect that there may be room for improvement.

Here we shall make a trial in this direction.
After reviewing the progress in the BLG theory and the ABJM theory, we shall turn to discuss a rather bold generalization by supersymmetrizing the three-algebra.

\section{Three-algebra}\label{BLG}

\begin{figure}[!ht]
\centering\includegraphics{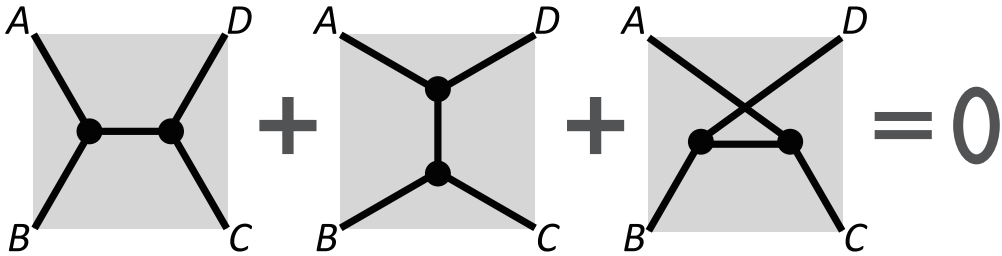}
\caption{A pictorial representation of the Jacobi identity in the Lie algebra.}
\label{jacobi}
\end{figure}

This section is devoted to a review of the three-algebra mainly from the algebraic viewpoint.
We shall start with Lie algebra, which is defined as a linear space equipped with a binary operation $[*,*]:{\mathfrak G}\times{\mathfrak G}\to{\mathfrak G}$, satisfying the following properties:
\begin{itemize}
\item antisymmetry,
$[X,Y]=-[Y,X]$,
\item bilinearity,
$[aX+bY,Z]=a[X,Z]+b[Y,Z]$, and
\item Jacobi identity,
$[[X,Y],Z]+[[Y,Z],X]+[[Z,X],Y]=0$.
\end{itemize}
Note that the Jacobi identity can be alternatively rewritten as
\begin{align}
[Z,[X,Y]]=[[Z,X],Y]+[X,[Z,Y]],
\label{Leibniz}
\end{align}
which takes the same form as the Leibniz rule, $D_Z(X\cdot Y)=(D_ZX)\cdot Y+X\cdot (D_ZY)$, if we regard the bracket product simultaneously as a derivative $D_Z(*)=[Z,*]$ and a product $X\cdot Y=[X,Y]$.
Using the structure constant defined as the coefficient of the commutators,
\begin{align}
[T^A,T^B]=f^{AB}{}_CT^C,
\end{align}
the Jacobi identity can be recast into
\begin{align}
f^{AB}{}_Zf^{ZC}{}_D+f^{BC}{}_Zf^{ZA}{}_D+f^{CA}{}_Zf^{ZB}{}_D=0,
\label{stu}
\end{align}
in the language of the structure constant.
For a mnemonic reason (and maybe a more profound one), we note that this identity can be regarded as the condition that the sum of the amplitudes in the $s$-channel, $t$-channel, and $u$-channel is vanishing
(see Figure \ref{jacobi}).
If contracted with the Killing metric $g^{ZC}$, the structure constant
\begin{align}
f^{ABC}=f^{AB}{}_Zg^{ZC}
\end{align}
is completely antisymmetric in exchanging the three indices.

\begin{figure}[!ht]
\centering\includegraphics{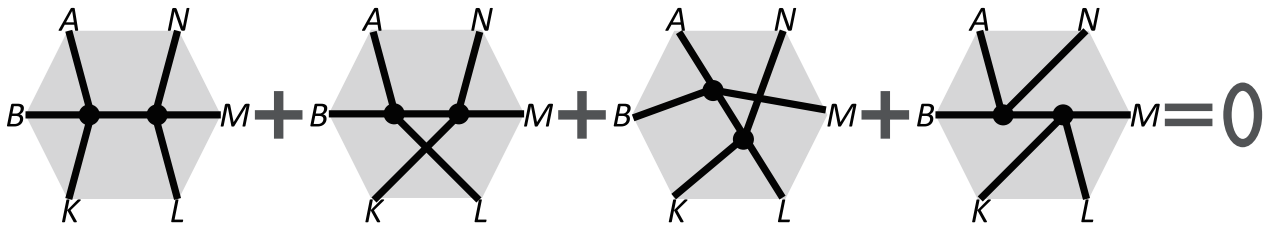}
\caption{A pictorial representation of the fundamental identity in the Nambu algebra.}
\label{fundamental}
\end{figure}

The Nambu algebra is a natural generalization of the Lie algebra.
Namely, it is the linear space equipped with a ternary operation $[*,*,*]:{\mathfrak G}\times{\mathfrak G}\times{\mathfrak G}\to{\mathfrak G}$, which satisfies the fundamental identity, a natural generalization of \eqref{Leibniz},
\begin{align}
&[T^A,T^B,[T^L,T^M,T^N]]\nonumber\\
&\quad=[[T^A,T^B,T^L],T^M,T^N]
+[T^L,[T^A,T^B,T^M],T^N]
+[T^L,T^M,[T^A,T^B,T^N]],
\end{align}
as well as the complete antisymmetry and trilinearity.
In terms of the structure constants
\begin{align}
[T^A,T^B,T^C]=F^{ABC}{}_DT^D,
\end{align}
the fundamental identity is given by
\begin{align}
F^{ABK}{}_ZF^{ZLMN}+F^{ABL}{}_ZF^{KZMN}
+F^{ABM}{}_ZF^{KLZN}+F^{ABN}{}_ZF^{KLMZ}=0.
\label{FI}
\end{align}
As before, we give a pictorial representation of the fundamental identity in Figure \ref{fundamental}.
Using the fundamental identity of the Nambu algebra, we can prove the gauge symmetry and the ${\cal N}=8$ supersymmetry \cite{BLsym}.
In this sense, we expect that the M2-branes in M-theory are described by the BLG theory.\footnote{Some other interesting ideas of studying M-theory using the Nambu bracket can be found, e.g., in Refs.~\cite{ALMY,PS,LP}.}

Since we expect the rank of the gauge group to be identified with the number of M2-branes, we are interested in finite-dimensional algebras.
However, if we require conditions of finite dimension, complete antisymmetry, and positive-definite metric, it can be proved \cite{Pap,GG} that the only nontrivial structure constant is
\begin{align}
F^{ABCD}=\varepsilon^{ABCD}.
\end{align}
The basic strategy in progress is to avoid these assumptions.
One fundamental example with infinite dimension is \eqref{NB}, which is the original Nambu bracket introduced in Ref.~\cite{Nambu} and was later interpreted as the M5-brane in Refs.~\cite{BLcomment,HoMa,HIMS,PSST}.
If we avoid the complete antisymmetry, this is nothing but the reformulation \cite{BLN6} of the ABJM theory \cite{ABJM} that we will discuss later.

Another interesting possibility is to avoid the positive-definite Killing metric \cite{GMR,HIM,GRGVRV}, which leads to the so-called Lorentzian BLG theory.
Namely, besides the Lie algebra generators, we add ``outer automorphism'' $u$ and ``center'' $v$:
\begin{align}
\{T^{A=1,\cdots,D},u,v\},
\end{align}
and consider the algebra
\begin{align}
[u,T^A,T^B]=f^{AB}{}_CT^C,\quad
[v,T^A,T^B]=0,\quad
[T^A,T^B,T^C]=-f^{ABC}v,
\label{uvLorentz}
\end{align}
equipped with the Lorentzian Killing metric
\begin{align}
g_{BA}T^AT^B+2uv.
\end{align}
Namely, the structure constant of the three-algebra is expressed in terms of that of the Lie algebra as
\begin{align}
F^{uABC}=-F^{ABC}{}_v=f^{ABC},\quad
F^{vABC}=-F^{ABC}{}_u=0,
\end{align}
and the fundamental identity follows from the properties of the Lie algebra.
To avoid the quantum-mechanical instability coming from the negative norm, a novel Higgs mechanism was proposed \cite{MuPa} by assigning a vacuum expectation value.

\section{ABJM theory}

The subsequent developments do not directly follow the idea of the Nambu 
bracket.
Initiated by the idea \cite{S} of supersymmetrizing the 2+1D non-Abelian Chern-Simons gauge theory, it was finally found that the system of $N$ coincident M2-branes on a geometry ${\mathbb C}^4/{\mathbb Z}_k$ is 
described by the ${\cal N}=6$ supersymmetric Chern-Simons theory \cite{ABJM}.
The gauge group is U$(N)_k\times$U$(N)_{-k}$ where the subscripts $k,-k$ stand for the Chern-Simons levels.
Besides the gauge field in the adjoint representation, we also need two pairs of bifundamental matters $(N,\overline N)$ and $(\overline N,N)$.

After various investigations, it turns out that the ABJM theory is correct.
This, however, does not directly mean that there is no room for improvement.
For example, although the supersymmetry preserved by the M2-branes on the flat spacetime is ${\cal N}=8$, only ${\cal N}=6$ is realized in the ABJM theory.
We shall look at the partition function and one-point functions of the half-BPS Wilson loop operator in the ABJM theory more carefully for a possible clue.

The partition function and one-point functions of the half-BPS Wilson loop on $S^3$ were calculated.
Using the localization technique, it was found that the expectation value originally defined by the infinite-dimensional path integral,
\begin{align}
\langle W_Y\rangle_k(N)
=\int DA_\mu\cdots e^{-S_\text{ABJM}[A_\mu,\cdots]}
\tr{\text P}\exp\int A_\mu dx^\mu+\cdots,
\end{align}
reduces to a finite-dimensional matrix integration \cite{KWY},
\begin{align}
\langle W_Y\rangle_k(N)
=\int_{{\mathbb R}^{2N}}
\frac{D^{N}\mu}{N!}\frac{D^{N}\nu}{N!}
W_Y(e^\mu,e^\nu)
\biggl(\frac{\prod_{a<b}^{N}2\sinh\frac{\mu_a-\mu_b}{2}
\prod_{c<d}^{N}2\sinh\frac{\nu_c-\nu_d}{2}}
{\prod_{a=1}^{N}\prod_{c=1}^{N}2\cosh\frac{\mu_a-\nu_c}{2}}\biggr)^2,
\label{mm}
\end{align}
where the integrations are
\begin{align}
D\mu_a=\frac{d\mu_a}{2\pi}e^{\frac{ik}{4\pi}\mu_a^2},\quad
D\nu_c=\frac{d\nu_c}{2\pi}e^{-\frac{ik}{4\pi}\nu_c^2},
\end{align}
and $W_Y(e^\mu,e^\nu)$ is the character coming from the Wilson loop in the representation $Y$.

If we look at the expression more carefully, it is not difficult to observe a deep relation to the supergroup U$(N|N)$ \cite{DT}.\footnote{The possible connection to the supergroup has already been observed in Ref.~\cite{GW} from the gauge group and the matter contents.}
Namely, the character $W_Y(e^\mu,e^\nu)$ appearing after the localization of the Wilson loop operator is nothing but that of the supergroup U$(N|N)$; the exponent appearing in the integrations,
\begin{align}
\frac{ik}{4\pi}\biggl(\sum_{a=1}^{N}\mu_a^2-\sum_{c=1}^{N}\nu_c^2\biggr),
\end{align}
takes the form of the supertrace of U$(N|N)$; and the integration measure can be regarded as the hyperbolic deformation of the invariant measure of the supergroup U$(N|N)$:
\begin{align}
\frac{\prod_{a<b}^{N}(\mu_a-\mu_b)\prod_{c<d}^{N}(\nu_c-\nu_d)}
{\prod_{a=1}^{N}\prod_{c=1}^{N}(\mu_a+\nu_c)}.
\end{align}
All of this evidence indicates that the supergroup structure is hidden behind the theory.

Hence, on one hand, the BLG theory is unsatisfactory because of its lack of examples of the three-algebra, while, on the other hand, the ABJM theory does not realize all of the supersymmetries ${\cal N}=8$ and the study of the partition function and one-point functions suggests a hidden supersymmetric gauge group.
Putting these considerations together, it is then interesting to ask whether we can take the idea of the super gauge group seriously by considering the supersymmetric generalization of the Nambu algebra.
Namely, we try to take a stance that the hidden super gauge group U$(N|N)$ appears not accidentally after the localization but as a gauge group for the BLG theory from the beginning.
It is only after we partially fix the fermionic gauge symmetry that we end up with the conventional ABJM theory with the bosonic gauge group U$(N)\times$U$(N)$.

It is known that in general the supergroup contains negative norms.
Typically, this is unacceptable in the path integral.
However, in the novel Higgs mechanism \cite{MuPa} we have already accepted one negative component.
Here let us simply assume that the multi-components of negative norms are still harmless.
In fact, in the matrix model \eqref{mm} the negative norms contribute as the Fresnel integral instead of the Gaussian integral.
Also, if we supersymmetrize the Nambu bracket as a gauge group, it is important to understand the role of the Chern-Simons level $k$ in this theory.
Let us neglect these problems and proceed for now.
It is important to come back to these problems in the future.

\section{Super-three-algebra}

We shall pose the question of whether it is possible to construct a Nambu superalgebra respecting the symmetry u$(N|N)$.
Namely, given the generators and the Killing metric of u$(N|N)$, we are interested in constructing a Nambu superalgebra satisfying the fundamental identity and other properties.\footnote{The classification of the super-three-algebra in the ${\cal N}=6$ setup \cite{BLN6} was done in Refs.~\cite{Pal,DMFOFME,CK}.
We are grateful to Takuya Matsumoto for pointing out these references to us.}

\begin{table}[!t]
\begin{align*}
f^{L^{\alpha_1}{}_{\alpha_2}L^{\beta_1}{}_{\beta_2}
L^{\gamma_1}{}_{\gamma_2}}
&=\delta^{\alpha_1}_{\gamma_2}\delta^{\beta_1}_{\alpha_2}
\delta^{\gamma_1}_{\beta_2}
-\delta^{\alpha_1}_{\beta_2}\delta^{\beta_1}_{\gamma_2}
\delta^{\gamma_1}_{\alpha_2},
&
f^{\widehat L^{\alpha_1}{}_{\alpha_2}\widehat L^{\beta_1}{}_{\beta_2}
\widehat L^{\gamma_1}{}_{\gamma_2}}
&=\delta^{\alpha_1}_{\gamma_2}\delta^{\beta_1}_{\alpha_2}
\delta^{\gamma_1}_{\beta_2}
-\delta^{\alpha_1}_{\beta_2}\delta^{\beta_1}_{\gamma_2}
\delta^{\gamma_1}_{\alpha_2},
\\
f^{R^{a_1}{}_{a_2}R^{b_1}{}_{b_2}R^{c_1}{}_{c_2}}
&=-\delta^{a_1}_{c_2}\delta^{b_1}_{a_2}\delta^{c_1}_{b_2}
+\delta^{a_1}_{b_2}\delta^{b_1}_{c_2}\delta^{c_1}_{a_2},
&
f^{\widehat R^{a_1}{}_{a_2}
\widehat R^{b_1}{}_{b_2}
\widehat R^{c_1}{}_{c_2}}
&=-\delta^{a_1}_{c_2}\delta^{b_1}_{a_2}\delta^{c_1}_{b_2}
+\delta^{a_1}_{b_2}\delta^{b_1}_{c_2}\delta^{c_1}_{a_2},
\\
f^{L^{\alpha_1}{}_{\alpha_2}Q^{\beta_1}{}_{b_2}S^{c_1}{}_{\gamma_2}}
&=\delta^{\alpha_1}_{\gamma_2}\delta^{\beta_1}_{\alpha_2}
\delta^{c_1}_{b_2},
&
f^{\widehat L^{\alpha_1}{}_{\alpha_2}
Q^{\beta_1}{}_{b_2}S^{c_1}{}_{\gamma_2}}
&=\delta^{\alpha_1}_{\gamma_2}\delta^{\beta_1}_{\alpha_2}
\delta^{c_1}_{b_2}
-\delta^{\alpha_1}_{\alpha_2}
\delta^{\beta_1}_{\gamma_2}\delta^{c_1}_{b_2}/N,
\\
f^{R^{a_1}{}_{a_2}Q^{\beta_1}{}_{b_2}S^{c_1}{}_{\gamma_2}}
&=-\delta^{a_1}_{b_2}\delta^{\beta_1}_{\gamma_2}\delta^{c_1}_{a_2},
&
f^{\widehat R^{a_1}{}_{a_2}Q^{\beta_1}{}_{b_2}S^{c_1}{}_{\gamma_2}}
&=-\delta^{a_1}_{b_2}\delta^{\beta_1}_{\gamma_2}\delta^{c_1}_{a_2}
+\delta^{a_1}_{a_2}
\delta^{\beta_1}_{\gamma_2}\delta^{c_1}_{b_2}/N.\\
&{\rm (a)}&&{\rm (b)}
\end{align*}
\caption{Structure constants of u$(N|N)$ (a) and psu$(N|N)$ (b).}
\label{lie}
\end{table}
Before proceeding to studying this question, we shall explain the superalgebra u$(N|N)$ and note that this algebra has a structure very similar to the Lorentzian three-algebra.
The superalgebra u$(N|N)$ contains two u$(N)$ factors as its bosonic subgroup, whose generators are denoted by $L^\alpha{}_\beta$ $(\alpha,\beta=1,2,\cdots,N)$ and $R^a{}_b$ $(a,b=1,2,\cdots,N)$, and fermionic generators connecting the two u$(N)$ factors denoted by $Q^c{}_\delta$ and $S^\gamma{}_d$.
The commutation relations are
\begin{align}
[L^\alpha{}_\beta,L^\gamma{}_\delta]
&=\delta^\gamma_\beta L^\alpha{}_\delta
-\delta^\alpha_\delta L^\gamma{}_\beta,&
[L^\alpha{}_\beta,Q^\gamma{}_d]&=\delta^\gamma_\beta Q^\alpha{}_d,&
[L^\alpha{}_\beta,S^c{}_\delta]&=-\delta^\alpha_\delta S^c{}_\beta,
\nonumber\\
[R^a{}_b,R^c{}_d]&=\delta^c_bR^a{}_d-\delta^a_dR^c{}_b,&
[R^a{}_b,Q^\gamma{}_d]&=-\delta^a_dQ^\gamma{}_b,&
[R^a{}_b,S^c{}_\delta]&=\delta^c_bS^a{}_\delta,
\nonumber\\
\{Q^\alpha{}_b,S^c{}_\delta\}&=\delta^c_bL^\alpha{}_\delta
+\delta^\alpha_\delta R^c{}_b,
\label{unn}
\end{align}
and the Killing metric is encoded in the quadratic Casimir operator:
\begin{align}
T=L^\alpha{}_\beta L^\beta{}_\alpha-R^a{}_bR^b{}_a
+S^a{}_\beta Q^\beta{}_a-Q^\alpha{}_bS^b{}_\alpha.
\end{align}
If we separate the trace part,
\begin{align}
u=\frac{L^\gamma{}_\gamma-R^c{}_c}{2\sqrt{N}},\quad
v=\frac{L^\gamma{}_\gamma+R^c{}_c}{\sqrt{N}},
\end{align}
and redefine the generators by subtracting the trace part,
\begin{align}
\widehat L^\alpha{}_\beta
=L^\alpha{}_\beta-\frac{\delta^\alpha_\beta}{N}L^\gamma{}_\gamma,\quad
\widehat R^a{}_b
=R^a{}_b-\frac{\delta^a_b}{N}R^c{}_c,
\end{align}
the algebra of the traceless part becomes
\begin{align}
[\widehat L^\alpha{}_\beta,\widehat L^\gamma{}_\delta]
&=\delta^\gamma_\beta\widehat L^\alpha{}_\delta
-\delta^\alpha_\delta\widehat L^\gamma{}_\beta,&
[\widehat L^\alpha{}_\beta,Q^\gamma{}_d]
&=\delta^\gamma_\beta Q^\alpha{}_d
-\frac{\delta^\alpha_\beta}{N}Q^\gamma{}_d,&
[\widehat L^\alpha{}_\beta,S^c{}_\delta]
&=-\delta^\alpha_\delta S^c{}_\beta
+\frac{\delta^\alpha_\beta}{N}S^c{}_\delta,
\nonumber\\
[\widehat R^a{}_b,\widehat R^c{}_d]
&=\delta^c_b\widehat R^a{}_d-\delta^a_d\widehat R^c{}_b,&
[\widehat R^a{}_b,Q^\gamma{}_d]&=-\delta^a_dQ^\gamma{}_b
+\frac{\delta^a_b}{N}Q^\gamma{}_d,&
[\widehat R^a{}_b,S^c{}_\delta]&=\delta^c_bS^a{}_\delta
-\frac{\delta^a_b}{N}S^c{}_\delta.
\end{align}
The nontrivial commutators related to the generator $u$ are
\begin{align}
[u,Q^\gamma{}_d]=Q^\gamma{}_d/\sqrt{N},\quad
[u,S^c{}_\delta]=-S^c{}_\delta/\sqrt{N},
\end{align}
while those related to the generator $v$ are
\begin{align}
\{Q^\alpha{}_b,S^c{}_\delta\}&=\delta^c_b\widehat L^\alpha{}_\delta
+\delta^\alpha_\delta\widehat R^c{}_b
+\delta^c_b\delta^\alpha_\delta v/\sqrt{N},
\end{align}
with $[v,*]=0$.
In this sense $u$ is an outer automorphism and $v$ is a center of the algebra.
This structure is often denoted as
\begin{align}
\text{u}(N|N)=\{u\}\ltimes\text{psu}(N|N)\ltimes\{v\},
\label{ltimes}
\end{align}
which basically means that $u$ only appears on the left-hand side of the 
nontrivial commutation relations, while $v$ only appears on the right-hand side.
Also, the quadratic Casimir operator encoding the Killing metric becomes
\begin{align}
T=\widehat L^\alpha{}_\beta\widehat L^\beta{}_\alpha
-\widehat R^a{}_b\widehat R^b{}_a
+S^a{}_\beta Q^\beta{}_a-Q^\alpha{}_bS^b{}_\alpha+2uv.
\end{align}
With these structures we can drop two generators $u$ and $v$ and define the algebra psu$(N|N)$.
All of these structures relate to the Lorentzian three-algebra.
This is why we have referred to $u$ and $v$ as outer automorphism and center in Sect.~\ref{BLG}.
To summarize, in terms of the structure constants, the superalgebras u$(N|N)$ and psu$(N|N)$ are given as in Table \ref{lie}.
Note that the Jacobi identity \eqref{stu} should be modified by
\begin{align}
f^{AB}{}_Zf^{ZC}{}_D+(-1)^{(B+C)A}f^{BC}{}_Zf^{ZA}{}_D
+(-1)^{C(A+B)}f^{CA}{}_Zf^{ZB}{}_D=0,
\label{superstu}
\end{align}
for the super-Lie algebra.
We can check explicitly that the structure constants of u$(N|N)$ and psu$(N|N)$ satisfy \eqref{superstu}.

\begin{table}[!t]
\begin{align*}
F^{L^{\alpha_1}{}_{\alpha_2}L^{\beta_1}{}_{\beta_2}
L^{\gamma_1}{}_{\gamma_2}L^{\delta_1}{}_{\delta_2}}
&=\#\bigl[\delta^{\alpha_1}_{\alpha_2}
(\delta^{\beta_1}_{\gamma_2}\delta^{\gamma_1}_{\delta_2}
\delta^{\delta_1}_{\beta_2}
-\delta^{\gamma_1}_{\beta_2}\delta^{\delta_1}_{\gamma_2}
\delta^{\beta_1}_{\delta_2})
-\delta^{\beta_1}_{\beta_2}
(\delta^{\gamma_1}_{\delta_2}\delta^{\delta_1}_{\alpha_2}
\delta^{\alpha_1}_{\gamma_2}
-\delta^{\delta_1}_{\gamma_2}\delta^{\alpha_1}_{\delta_2}
\delta^{\gamma_1}_{\alpha_2})\\
&\qquad+\delta^{\gamma_1}_{\gamma_2}
(\delta^{\delta_1}_{\alpha_2}\delta^{\alpha_1}_{\beta_2}
\delta^{\beta_1}_{\delta_2}
-\delta^{\alpha_1}_{\delta_2}\delta^{\beta_1}_{\alpha_2}
\delta^{\delta_1}_{\beta_2})
-\delta^{\delta_1}_{\delta_2}
(\delta^{\alpha_1}_{\beta_2}\delta^{\beta_1}_{\gamma_2}
\delta^{\gamma_1}_{\alpha_2}
-\delta^{\beta_1}_{\alpha_2}\delta^{\gamma_1}_{\beta_2}
\delta^{\alpha_1}_{\gamma_2})\bigr],
\\
F^{L^{\alpha_1}{}_{\alpha_2}L^{\beta_1}{}_{\beta_2}
L^{\gamma_1}{}_{\gamma_2}R^{d_1}{}_{d_2}}
&=\#\bigl[\delta^{d_1}_{d_2}
(\delta^{\alpha_1}_{\beta_2}\delta^{\beta_1}_{\gamma_2}
\delta^{\gamma_1}_{\alpha_2}
-\delta^{\beta_1}_{\alpha_2}\delta^{\gamma_1}_{\beta_2}
\delta^{\alpha_1}_{\gamma_2})\bigr],
\\
F^{L^{\alpha_1}{}_{\alpha_2}
R^{b_1}{}_{b_2}R^{c_1}{}_{c_2}R^{d_1}{}_{d_2}}
&=\#\bigl[\delta^{\alpha_1}_{\alpha_2}
(\delta^{b_1}_{c_2}\delta^{c_1}_{d_2}\delta^{d_1}_{b_2}
-\delta^{c_1}_{b_2}\delta^{d_1}_{c_2}\delta^{b_1}_{d_2})\bigr],
\\
F^{R^{a_1}{}_{a_2}R^{b_1}{}_{b_2}R^{c_1}{}_{c_2}R^{d_1}{}_{d_2}}
&=\#\bigl[\delta^{a_1}_{a_2}
(\delta^{b_1}_{c_2}\delta^{c_1}_{d_2}\delta^{d_1}_{b_2}
-\delta^{c_1}_{b_2}\delta^{d_1}_{c_2}\delta^{b_1}_{d_2})
-\delta^{b_1}_{b_2}
(\delta^{c_1}_{d_2}\delta^{d_1}_{a_2}\delta^{a_1}_{c_2}
-\delta^{d_1}_{c_2}\delta^{a_1}_{d_2}\delta^{c_1}_{a_2})\\
&\qquad+\delta^{c_1}_{c_2}
(\delta^{d_1}_{a_2}\delta^{a_1}_{b_2}\delta^{b_1}_{d_2}
-\delta^{a_1}_{d_2}\delta^{b_1}_{a_2}\delta^{d_1}_{b_2})
-\delta^{d_1}_{d_2}
(\delta^{a_1}_{b_2}\delta^{b_1}_{c_2}\delta^{c_1}_{a_2}
-\delta^{b_1}_{a_2}\delta^{c_1}_{b_2}\delta^{a_1}_{c_2})\bigr],
\\
F^{L^{\alpha_1}{}_{\alpha_2}L^{\beta_1}{}_{\beta_2}
Q^{\gamma_1}{}_{c_2}S^{d_1}{}_{\delta_2}}
&=\#\bigl[\delta^{d_1}_{c_2}
(\delta^{\alpha_1}_{\alpha_2}\delta^{\beta_1}_{\delta_2}
\delta^{\gamma_1}_{\beta_2}
-\delta^{\alpha_1}_{\delta_2}\delta^{\beta_1}_{\beta_2}
\delta^{\gamma_1}_{\alpha_2})\bigr]
+\#\bigl[\delta^{d_1}_{c_2}
(\delta^{\alpha_1}_{\beta_2}\delta^{\beta_1}_{\delta_2}
\delta^{\gamma_1}_{\alpha_2}
-\delta^{\alpha_1}_{\delta_2}\delta^{\beta_1}_{\alpha_2}
\delta^{\gamma_1}_{\beta_2})\bigr],
\\
F^{L^{\alpha_1}{}_{\alpha_2}R^{b_1}{}_{b_2}
Q^{\gamma_1}{}_{c_2}S^{d_1}{}_{\delta_2}}
&=\#\bigl[\delta^{\alpha_1}_{\alpha_2}\delta^{b_1}_{b_2}
\delta^{\gamma_1}_{\delta_2}\delta^{d_1}_{c_2}\bigr]
+\#\bigl[\delta^{\alpha_1}_{\alpha_2}\delta^{b_1}_{c_2}
\delta^{\gamma_1}_{\delta_2}\delta^{d_1}_{b_2}\bigr]
+\#\bigl[\delta^{\alpha_1}_{\delta_2}\delta^{b_1}_{b_2}
\delta^{\gamma_1}_{\alpha_2}\delta^{d_1}_{c_2}\bigr]\\
&\qquad+\#\bigl[\delta^{\alpha_1}_{\delta_2}\delta^{b_1}_{c_2}
\delta^{\gamma_1}_{\alpha_2}\delta^{d_1}_{b_2}\bigr],
\\
F^{R^{a_1}{}_{a_2}R^{b_1}{}_{b_2}
Q^{\gamma_1}{}_{c_2}S^{d_1}{}_{\delta_2}}
&=\#\bigl[\delta^{\gamma_1}_{\delta_2}
(\delta^{a_1}_{a_2}\delta^{b_1}_{c_2}\delta^{d_1}_{b_2}
-\delta^{a_1}_{c_2}\delta^{b_1}_{b_2}\delta^{d_1}_{a_2})\bigr]
+\#\bigl[\delta^{\gamma_1}_{\delta_2}
(\delta^{a_1}_{b_2}\delta^{b_1}_{c_2}\delta^{d_1}_{a_2}
-\delta^{a_1}_{c_2}\delta^{b_1}_{a_2}\delta^{d_1}_{b_2})\bigr],
\\
F^{Q^{\alpha_1}{}_{a_2}Q^{\beta_1}{}_{b_2}
S^{c_1}{}_{\gamma_2}S^{d_1}{}_{\delta_2}}
&=\#\bigl[\delta^{\alpha_1}_{\gamma_2}\delta^{\beta_1}_{\delta_2}
\delta^{c_1}_{a_2}\delta^{d_1}_{b_2}
+\delta^{\alpha_1}_{\delta_2}\delta^{\beta_1}_{\gamma_2}
\delta^{c_1}_{b_2}\delta^{d_1}_{a_2}\bigr]
+\#\bigl[\delta^{\alpha_1}_{\gamma_2}\delta^{\beta_1}_{\delta_2}
\delta^{c_1}_{b_2}\delta^{d_1}_{a_2}
+\delta^{\alpha_1}_{\delta_2}\delta^{\beta_1}_{\gamma_2}
\delta^{c_1}_{a_2}\delta^{d_1}_{b_2}\bigr].
\end{align*}
\caption{An ansatz for the super-three-algebra constructed based on the 
generators and the Killing metric of u$(N|N)$.}
\label{ansatz}
\end{table}

Now let us generalize the super-Lie algebra to the three-algebra. 
Namely, we write down the structure constants given in Table \ref{ansatz}, which are consistent with the tensorial structure of u$(N)\times$u$(N)$ and the complete antisymmetry.
After that, to see what coefficients are allowed, we substitute the ansatz into the fundamental identity \eqref{FI}, which, for the super case, should be modified into
\begin{align}
&F^{ABK}{}_ZF^{ZLMN}
+(-1)^{(A+B)K}F^{ABL}{}_ZF^{KZMN}\nonumber\\
&\quad+(-1)^{(A+B)(K+L)}F^{ABM}{}_ZF^{KLZN}
+(-1)^{(A+B)(K+L+M)}F^{ABN}{}_ZF^{KLMZ}=0.
\end{align}
In this way, we have obtained a set of equations for the coefficients.
Up to a sign ambiguity (which should only relate to the conventions), the 
solution is given as in Table \ref{answer}.

\begin{table}[!t]
\begin{align*}
&\sqrt{N}F^{L^{\alpha_1}{}_{\alpha_2}L^{\beta_1}{}_{\beta_2}
L^{\gamma_1}{}_{\gamma_2}L^{\delta_1}{}_{\delta_2}}
\nonumber\\
&=\delta^{\alpha_1}_{\alpha_2}
f^{\widehat L^{\beta_1}{}_{\beta_2}
\widehat L^{\gamma_1}{}_{\gamma_2}
\widehat L^{\delta_1}{}_{\delta_2}}
-\delta^{\beta_1}_{\beta_2}
f^{\widehat L^{\gamma_1}{}_{\gamma_2}
\widehat L^{\delta_1}{}_{\delta_2}
\widehat L^{\alpha_1}{}_{\alpha_2}}
+\delta^{\gamma_1}_{\gamma_2}
f^{\widehat L^{\delta_1}{}_{\delta_2}
\widehat L^{\alpha_1}{}_{\alpha_2}
\widehat L^{\beta_1}{}_{\beta_2}}
-\delta^{\delta_1}_{\delta_2}
f^{\widehat L^{\alpha_1}{}_{\alpha_2}
\widehat L^{\beta_1}{}_{\beta_2}
\widehat L^{\gamma_1}{}_{\gamma_2}},
\\
&\sqrt{N}F^{L^{\alpha_1}{}_{\alpha_2}L^{\beta_1}{}_{\beta_2}
L^{\gamma_1}{}_{\gamma_2}R^{d_1}{}_{d_2}}
=\delta^{d_1}_{d_2}
f^{\widehat L^{\alpha_1}{}_{\alpha_2}
\widehat L^{\beta_1}{}_{\beta_2}
\widehat L^{\gamma_1}{}_{\gamma_2}},
\\
&\sqrt{N}F^{L^{\alpha_1}{}_{\alpha_2}
R^{b_1}{}_{b_2}R^{c_1}{}_{c_2}R^{d_1}{}_{d_2}}
=\delta^{\alpha_1}_{\alpha_2}
f^{\widehat R^{b_1}{}_{b_2}
\widehat R^{c_1}{}_{c_2}
\widehat R^{d_1}{}_{d_2}},
\\
&\sqrt{N}F^{R^{a_1}{}_{a_2}R^{b_1}{}_{b_2}R^{c_1}{}_{c_2}R^{d_1}{}_{d_2}}
\nonumber\\
&=-\delta^{a_1}_{a_2}
f^{\widehat R^{b_1}{}_{b_2}
\widehat R^{c_1}{}_{c_2}
\widehat R^{d_1}{}_{d_2}}
+\delta^{b_1}_{b_2}
f^{\widehat R^{c_1}{}_{c_2}
\widehat R^{d_1}{}_{d_2}
\widehat R^{a_1}{}_{a_2}}
-\delta^{c_1}_{c_2}
f^{\widehat R^{d_1}{}_{d_2}
\widehat R^{a_1}{}_{a_2}\widehat R^{b_1}{}_{b_2}}
+\delta^{d_1}_{d_2}
f^{\widehat R^{a_1}{}_{a_2}
\widehat R^{b_1}{}_{b_2}
\widehat R^{c_1}{}_{c_2}},
\\
&\sqrt{N}F^{L^{\alpha_1}{}_{\alpha_2}L^{\beta_1}{}_{\beta_2}
Q^{\gamma_1}{}_{c_2}S^{d_1}{}_{\delta_2}}
=\delta^{\alpha_1}_{\alpha_2}
f^{\widehat L^{\beta_1}{}_{\beta_2}
Q^{\gamma_1}{}_{c_2}S^{d_1}{}_{\delta_2}}
-\delta^{\beta_1}_{\beta_2}
f^{\widehat L^{\alpha_1}{}_{\alpha_2}
Q^{\gamma_1}{}_{c_2}S^{d_1}{}_{\delta_2}},
\\
&\sqrt{N}F^{L^{\alpha_1}{}_{\alpha_2}R^{b_1}{}_{b_2}
Q^{\gamma_1}{}_{c_2}S^{d_1}{}_{\delta_2}}
=\delta^{\alpha_1}_{\alpha_2}
f^{\widehat R^{b_1}{}_{b_2}Q^{\gamma_1}{}_{c_2}S^{d_1}{}_{\delta_2}}
+\delta^{b_1}_{b_2}
f^{\widehat L^{\alpha_1}{}_{\alpha_2}
Q^{\gamma_1}{}_{c_2}S^{d_1}{}_{\delta_2}},
\\
&\sqrt{N}F^{R^{a_1}{}_{a_2}R^{b_1}{}_{b_2}
Q^{\gamma_1}{}_{c_2}S^{d_1}{}_{\delta_2}}
=-\delta^{a_1}_{a_2}
f^{\widehat R^{b_1}{}_{b_2}Q^{\gamma_1}{}_{c_2}S^{d_1}{}_{\delta_2}}
+\delta^{b_1}_{b_2}
f^{\widehat R^{a_1}{}_{a_2}Q^{\gamma_1}{}_{c_2}S^{d_1}{}_{\delta_2}},
\\
&F^{Q^{\alpha_1}{}_{a_2}Q^{\beta_1}{}_{b_2}
S^{c_1}{}_{\gamma_2}S^{d_1}{}_{\delta_2}}
=0.
\end{align*}
\caption{A solution to the super-three-algebra constructed based on the 
generators and the Killing metric of u$(N|N)$.
The result is given in terms of the structure constants for psu$(N|N)$ given in 
Table \ref{lie}.}
\label{answer}
\end{table}

The result in Table \ref{answer} can be summarized as
\begin{align}
F^{ABCD}=\Delta^Af^{\widehat B\widehat C\widehat D}
-\Delta^Bf^{\widehat A\widehat C\widehat D}
+\Delta^Cf^{\widehat A\widehat B\widehat D}
-\Delta^Df^{\widehat A\widehat B\widehat C},
\end{align}
where $\Delta^A=\Delta^{G^A}$ and $\Delta_A=\Delta_{G^A}$ are defined as
\begin{align}
\Delta^{G^A}=\begin{cases}
\delta^{\alpha_1}_{\alpha_2}/\sqrt{N}&G^A=L^{\alpha_1}{}_{\alpha_2}\\
-\delta^{a_1}_{a_2}/\sqrt{N}&G^A=R^{a_1}{}_{a_2}\\
0&\mbox{otherwise}
\end{cases},\quad
\Delta_{G^A}=\begin{cases}
\delta^{\alpha_2}_{\alpha_1}/\sqrt{N}&G^A=L^{\alpha_1}{}_{\alpha_2}\\
\delta^{a_2}_{a_1}/\sqrt{N}&G^A=R^{a_1}{}_{a_2}\\
0&\mbox{otherwise}
\end{cases},
\label{deltaA}
\end{align}
while $f^{\widehat B\widehat C\widehat D}$ is the structure constants for the corresponding traceless generators of psu$(N|N)$ given in Table \ref{lie}.
In $f^{\widehat B\widehat C\widehat D}$ we identify the fermionic generators of u$(N|N)$ with those of psu$(N|N)$, $\widehat Q^\alpha{}_b=Q^\alpha{}_b, \widehat S^a{}_\beta=S^a{}_\beta$.
These quantities satisfy
\begin{align}
\Delta_Z\Delta^Z=0,\quad
\Delta_Zf^{\widehat A\widehat B\widehat Z}=0,
\label{DeltaZ}
\end{align}
where the former relation can be confirmed directly by \eqref{deltaA}, while the latter holds because the psu$(N|N)$ generators are traceless.
Then, we can check all cases of the fundamental identity directly.
In fact, due to \eqref{DeltaZ}, the only possible contributions come from $\Delta^X\Delta^Y$ with $X=A,B$ and $Y=K,L,M,N$ and $\Delta^{Y_1}\Delta^{Y_2}$ with $Y_1,Y_2=K,L,M,N$.
For the former case, the coefficients reduce to the Jacobi identity \eqref{superstu}, while the coefficients for the latter case vanish simply from the antisymmetry of the structure constants.
This algebra looks very similar to the Lorentzian three-algebra.
In fact, we can further identify them by computing
\begin{align}
F^{u\widehat A\widehat B\widehat C}
=-F^{\widehat A\widehat B\widehat C}{}_v
=f^{\widehat A\widehat B\widehat C},\quad
F^{v\widehat A\widehat B\widehat C}
=-F^{\widehat A\widehat B\widehat C}{}_u
=0,
\end{align}
which indicates that the role of the unknown Lie algebra used in defining the three-algebra \eqref{uvLorentz} is played by the traceless algebra psu$(N|N)$.

\section{Summary}

As we have seen here, Nambu brackets have played an important role in understanding the M2-branes.
In particular, recent progress in understanding the multiple M2-branes starts from the Nambu algebra.
From it we arrive at the BLG theory, and afterwards the ABJM theory.
The ABJM theory turns out to describe the multiple M2-branes correctly, though there may be room for improvement, because the supersymmetry is realized only up to ${\cal N}=6$.
From the expression of the partition function of the ABJM theory obtained from the localization, we expect that the supergroup U$(N|N)$ may play a role in the gauge symmetry.

Putting these considerations together, we are naturally led to the idea of supersymmetrizing the Nambu algebra.
We have found that the Nambu algebra built on the superalgebra u$(N|N)$ does exist.
The result is given explicitly in terms of the structure constants of psu$(N|N)$.
Although this is not surprising, this indicates that the framework of \eqref{ltimes} fits well with the Lorentzian three-algebra \eqref{uvLorentz}.

A natural question would be whether the super-three-algebra works for other superalgebras like u$(N_1|N_2)$ or osp$(N_1|2N_2)$, where the theories also enjoy an enhanced supersymmetry of ${\cal N}=6$ or ${\cal N}=5$ \cite{HLLLP2,ABJ}.
Many interesting relations between these theories have been found recently, such as chiral projection \cite{HMO1,MePu,Ho2,Ok2,MS2,MN5}, the Giambelli identity \cite{HHMO,MaMo}, the open-closed duality \cite{MM,HaOk}, quiver doubling \cite{HoMo,MS1}, and the quantum Wronskian relation \cite{GHM}.
It would be nice if we could study these relations from algebraic viewpoints.

To recapitulate, our stance is to assume the existence of a fundamental BLG theory with gauge group given by the super-three-algebra.
After partially fixing the fermionic gauge symmetry, we expect that the theory reduces to the original ABJM theory where the superalgebra is invisible.
The superalgebra appears again when computing the partition function or one-point functions of the half-BPS Wilson loop.
For this expectation to work, we need to construct a full theory by considering the superpartners of the BLG theory and study the partial gauge fixing of the fermionic symmetry or compute the partition function using the super-BLG theory.
One difficulty would be the negative norms in the quantization.
We expect, however, that they are harmless in the Fresnel integral.

In Ref.~\cite{HMS} the directions of the negative components were interpreted as the compactification directions.
Our interpretation seems different from this viewpoint.
It would be interesting to see the relation between the two interpretations.

\section*{Acknowledgements}

We are grateful to Pei-Ming Ho, Takuya Matsumoto, and Yutaka Matsuo for valuable comments on our manuscript and Koji Hashimoto, Shinji Hirano, Seiji Terashima, and Satoshi Yamaguchi for discussions.
We would like to thank the PTEP production team for their grammatical check on our manuscript.
The work of S.M.\ is supported by JSPS Grant-in-Aid for Scientific
Research (C) \# 26400245.
S.M.\ would like to thank Yukawa Institute for Theoretical Physics at Kyoto University for their hospitality.

\end{document}